\documentclass[a4paper,11pt]{article}

\usepackage{jinstpub}
\usepackage{graphicx}

\title{Development of a PET/EPRI combined imaging system for assessing tumor hypoxia}
    
\author[a]{H. Kim,\note{Corresponding author.}}
\author[b,c]{B. Epel}
\author[b,c]{S. Sundramoorthy }
\author[a]{H.-M. Tsai}
\author[b,c]{E. Barth}
\author[a,b,c]{I. Gertsenshteyn}
\author[b,c]{H. Halpern}
\author[d]{Y. Hua}
\author[e]{Q. Xie}
\author[a]{C.-T. Chen}
\author[a]{C.-M. Kao}

\affiliation[a]{Department of Radiology, University of Chicago, Chicago, IL 60637}
\affiliation[b]{Department of Radiation and Cellular Oncology, University of Chicago, Chicago, IL 60637}
\affiliation[c]{Center for EPR Imaging In Vivo Physiology, University of Chicago, Chicago, IL 60637}
\affiliation[d]{Raycan Technology Co, Ltd., Suzhou, Jiangsu, China}
\affiliation[e]{Huazhong University of Science and Technology, Biomedical Engineering Department, Wuhan, Hubei, China}

\emailAdd{heejongkim@uchicago.edu}

\abstract{
Precise quantitative delineation of tumor hypoxia is essential in radiation therapy treatment planning
to improve the treatment efficacy by targeting hypoxic sub-volumes.
We developed a combined imaging system of positron emission tomography (PET) and 
electron para-magnetic resonance imaging (EPRI) of molecular oxygen
to investigate the accuracy of PET imaging in assessing tumor hypoxia.
The PET/EPRI combined imaging system aims to use EPRI to precisely measure the oxygen partial pressure in tissues.
This will evaluate the validity of PET hypoxic tumor imaging 
by (near) simultaneously acquired EPRI as ground truth. 
The combined imaging system was constructed 
by integrating a small animal PET scanner (inner ring diameter 62 mm and axial field of view 25.6 mm)
and an EPRI subsystem (field strength 25 mT and resonant frequency 700 MHz).
The compatibility between the PET and EPRI subsystems were tested with both phantom and animal imaging.
Hypoxic imaging on a tumor mouse model using $^{18}$F-fluoromisonidazole radio-tracer
was conducted with the developed PET/EPRI system.
We report the development and initial imaging results obtained from the PET/EPRI combined imaging system.  
}

\keywords{Multi-modality systems, Positron Emission Tomography, Electron-Paramagnetic Resonance Imaging, Tumor Hypoxia}

\begin{document}
\maketitle
\flushbottom

\section{Introduction}
Because of the rapid proliferation of cancer cells outgrowing blood supply,
the amount of oxygen available is limited in some tumor regions distant from blood vessels.
Tumor hypoxia ~\citep{hockel_2001, walsh_2014} refers to the lower oxygen concentration in tumor tissue, 
typically below 10 torr,
and it is a characteristic feature of human and animal solid tumors.
Hypoxia is known to promote tumor progression and migration 
by changing its metabolism to adapt to oxygen deprived environment~\citep{sullivan_2007, lu_2010}.
It has been known that hypoxic tumors are more resistant to radiation therapy~\citep{gray_1953, brizel_1999, moeller_2007},
and require additional dose delivery for effective treatment.
Therefore, precise delineation of hypoxic tumor region is essential
for targeting only hypoxic sub-volume within the tumor in radiation therapy to improve the treatment outcome~\citep{rajendran_2006, thorwarth_2007, lee_2014, epel_2019}.

In the clinic, 
PET imaging with radio-tracers such as $^{18}$F-fluoromisonidazole (F-MISO)~\citep{rasey_1987, rajendran_2015}
has been used for hypoxic tumor imaging,
which exploits the mechanism of misonidazole binding intracellularly
at oxygen-depleted sites 2-4 hours post-injection. 
However, F-MISO PET imaging has not yet beneficially guided radiation therapy treatment for hypoxic tumors. 
For example, a recent phase-II clinical study \citep{vera_2017} showed that 
dose painting based on F-MISO PET imaging did not significantly improve the treatment outcomes.

Currently, the F-MISO uptake and binding  mechanisms in hypoxic tumor are not clearly understood~\citep{masaki_2015}.
Other factors, such as the vascular structure and pH in the micro-environment, 
have been observed to affect the F-MISO absorption within a hypoxic region~\citep{monnich_2011, kipp_2004}.
These factors substantially complicate the correlation between the degree of hypoxia and F-MISO uptake. 
On the other hand, in preclinical studies, 
the effectiveness of oxygen imaging by electron para-magnetic resonance imaging (EPRI)
for targeting hypoxic tumors has been demonstrated~\citep{elas_2013, epel_2017, epel_2019}.
EPRI is a non-invasive imaging technology 
based on the detection of unpaired electron spins of injectable spin probe subjected to a magnetic field, 
and is capable of quantitative measurements of the absolute partial pressure of oxygen (pO$_2$) in tissue~\citep{berliner_1985, halpern_1989, epel_2014}.
However, EPRI is not available for clinical use presently.

We recently initiated a project to investigate the potential of PET imaging 
in assessing tumor hypoxia in small-animal settings.
In the project, we developed a combined small-animal PET and EPRI imaging system 
to allow simultaneous or rapid succession imaging 
so that we can use EPR oxygen imaging as a gold standard for in vivo measurement of tissue pO$_2$
to evaluate the accuracy of PET hypoxia imaging, and to develop correction methods if necessary.
The (near) simultaneous PET/EPR operation ensures the recording of the same physiological
and biochemical changes in the tissue,
and accordingly reflects the correlations between PET and oxygen images more accurately.
PET-EPR image registration is not required with the combined system
which is not the case for separate PET and EPR scanners.

In this article, 
we present the development of the combined PET/EPRI system 
and the initial imaging results obtained by using the system.

\section{Material and Methods}
\begin{figure}[h!]
\includegraphics[height=4.0cm]{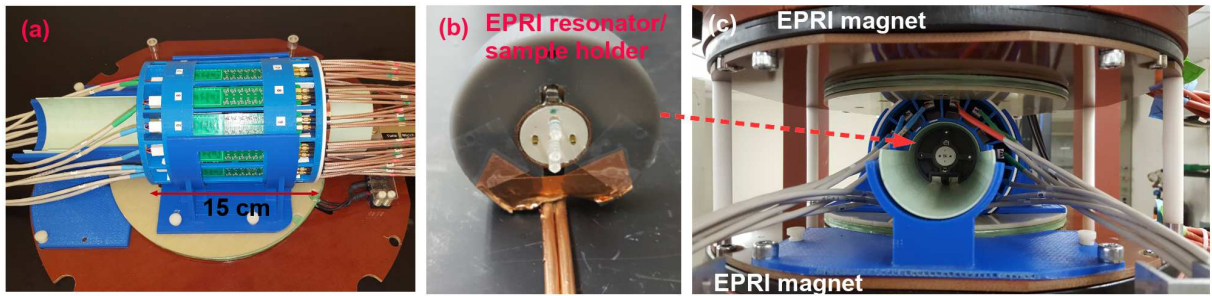} 
\caption{
  (a) The PET detector before integration.
  (b) A loop-gap type EPRI RF resonator. A phantom with capillary tubes is positioned at the center of the resonator for imaging.
  (c) A PET/EPRI combined system. The PET detector is installed between two permanent EPRI magnets.
}
\label{fig:1} 
\end{figure}
\subsection{PET subsystem}
The PET subsystem was originally developed as an insert for a small animal imaging MRI.
It consists of 14 detector modules,
which are installed within a cylindrical supporting structure
that has a 60 mm inner diameter and a 115 mm outer diameter.
Each detector module uses an array of 8x4 LYSO scintillators (each LYSO crystal is 3x3x10 mm$^3$) 
and two Hamamatsu MPPC arrays (S13361-3050-NE or S12642-0404-PA).
The scintillators within the array are optically isolated by using enhanced specular reflector (ESR, 3M$^{TM}$),
and are coupled individually to SiPMs (3.2 mm pitch) in the MPPC array.  
The axial field of view of the PET detector is 25.6 mm.
A strip-line based  multiplexing method~\citep{kim_2015} is used for SiPM signal readout.
An earlier prototype detector module~\citep{kim_2015, kim_2016}
used 8 SiPMs of 4.0x4.4 mm$^2$ active area (SPM42-75, STM) on a strip-line.
In comparison, the detector module of the reported system used Hammamatsu MPPC arrays instead
due to its smaller pixel size and more uniform gain between SiPM pixels.
It also has 16 SiPMs on a strip-line to achieve a higher multiplexing ratio.
The 32 SiPM outputs from a detector module are routed to two strip-lines 
implemented in a strip-line board (SLB), which is a 18x110 mm$^2$ printed circuit board.
The SLB outputs are connected to a multi-voltage threshold (MVT) waveform sampling board~\citep{xi_2013} 
placed away from the MRI magnetic and radio-frequency (RF) fields 
via 5 m long miniature coaxial cables for signal digitization.
The MVT board implements voltage-based sampling of PET signals as following. 
The board provides 4 user-definable voltage thresholds, 
which in this study are set to 50, 150, 300 and 600 mV. 
The leading and trailing transitions of a PET signal over and below these thresholds are obtained by comparators 
and the transition times are determined by time-to-digital converters (TDC) with 95 ps bin width. 
The known voltage thresholds and the TDC produced time stamps are packed along with the channel number 
to a data acquisition computer through an Ethernet interface. 
The comparators and TDCs are all implemented by using field programmable gate arrays (FPGAs).
The performance and MR compatibility of this prototypical PET insert was previously characterized 
in experiments with a Bruker BioSpec 9.4 T MR scanner (Billerica MA).
Figures~\ref{fig:1}(a) shows a photo of the PET insert. 
More details on the design and performance characterization of the PET insert are described in our previous publication\citep{kim_2020}.

\subsection{EPRI subsystem}
The EPRI subsystem is a 700 ($\pm$ 20) MHz pulse imager/relaxometer utilizing an inversion recovery electron spin echo (IRESE) acquisition method~\citep{epel_2014}.
It consists of a parallel face 25 mT permanent magnet with a 12 cm gap, 2 mT main field offset coil,
three planar orthogonal gradient coils (30 mT/m maximum gradient), an RF resonator and RF control electronics.
The resonator is located inside the PET system
while the offset coils are placed in between the PET system and the magnet.
The RF resonator of the EPRI, shown in Figure~\ref{fig:1}(b),
is a loop-gap resonator in reflection mode for both exciting and detecting RF signals at 700 MHz.
The resonator has a cylindrical 19 mm-diameter access that is 15mm in length. 
It is embedded in a plastic spacer for positioning at the isocenter of the gradient fields and the PET system.
The object to be imaged is placed within the resonator as shown in Figure~\ref{fig:1}(c).
A 4 kW TOMCO RF amplifier is used to power RF pulses,
though only 250 W was used for our experiments.
The homodyne bridge scheme and control method similar to~\citep{epel_2008} is used.
The instrument is controlled using SpecMan4EPR software~\citep{epel_2005}.

\subsection{Combined system and alignment}
Figure~\ref{fig:1}(c) shows the combined PET/EPRI system encased in a 40x40x31 cm$^3$ frame. 
Both the center of the resonator and that of the PET field of view are aligned with the center of the magnets. 

Data acquisition and processing of the PET and EPRI subsystems
are accomplished separately by two computers.
For EPRI, the IRESE imaging sequence was used with 201 projections
and a maximum gradient of 7.5 mT/m.
Nine time delays between the inversion and the electron spin echo detection
were used to measure relaxation times.
Image acquisition time was about 7 minutes.
The data were reconstructed using a filtered back-projection (FBP) algorithm~\citep{ahn_2007}. 
PET data were acquired as list-mode data of singles events; 
post acquisition they were processed to produce coincidence data that were then reconstructed 
by using a line-of-response based maximum likelihood expectation maximization (MLEM) algorithm~\citep{shepp_1982}.
The PET image reconstruction code was previously validated 
in imaging studies of a $^{22}$Na point source, a  $^{68}$Ge line source, and a custom-made resolution phantom~\citep{kim_2020}.

\begin{figure}[h!]
\begin{center}
\includegraphics[height=4.0cm]{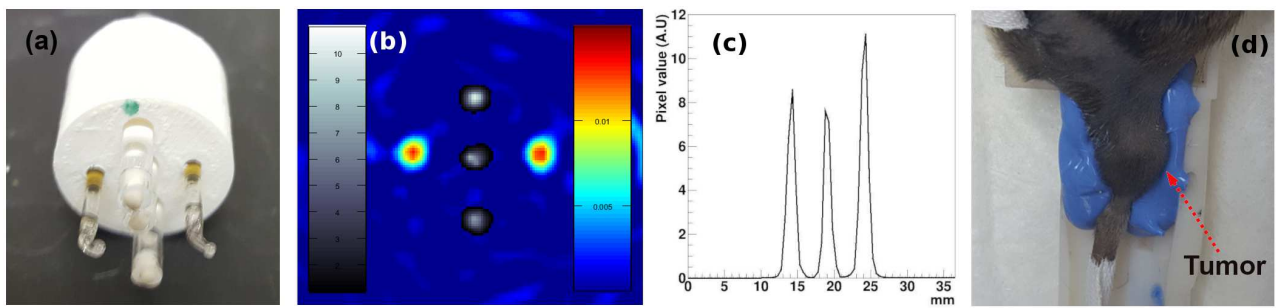}
\end{center}
\caption{
  (a) An imaging phantom with 3 capillary tubes filled with FDG and 2 tubes filled with trityl spin probe solution.
  (b) A PET/EPRI registered image of the imaging phantom.
  (c) A profile of the PET image 
  on the line through the centers of the 3 capillary tubes filled with FDG.
  (d) A tumor-borne mouse leg is immobilized in the animal bed using polyvinyl siloxane mold cast; 
  the bed is placed in the resonator for imaging.   
}
\label{fig:2} 
\end{figure}

Alignment between the PET and EPRI system was assessed by imaging a phantom
shown in Figure~\ref{fig:2}(a).
The phantom holds 5 capillary tubes;
three tubes were filled with  $^{18}$F-fluorodeoxyglucose (FDG) for PET,
and the other two tubes were filled with 1 mM of trityl spin probe solution ~\citep{kuzhelev_2015} for EPRI.
Figure~\ref{fig:2}(b) shows a PET/EPRI combined image of the phantom.
The PET image of the 3 FDG-filled capillary tubes is shown in greyscale,
and EPR image of the 2 trityl-filled capillary tubes is shown in color.
Figure~\ref{fig:2}(c) shows an intensity profile of the PET image through the center of the 3 capillary tubes.
The PET and EPRI systems were previously evaluated 
to have an image resolution of approximately 1.6 mm and 1.0 mm, respectively~\citep{kim_2020}. 
The intensity profiles (EPRI intensity profiles not shown) indicate 
that these resolutions are maintained by the combined system.

\subsection{Small animal imaging}
Small animal imaging followed the protocol approved
by the University of Chicago's Institutional Animal Care and Use Committee.
Squamous cell carcinoma tumor cells were grown in the leg of a C3H mouse.
The animal was positioned into an animal bed
which was equipped with a set of 4 non-parallel capillary tubes for registration purpose.
The tumor area was secured using polyvinyl siloxane (PVS) dental mold material, 
as shown in Figures~\ref{fig:2}(d), and the mold was placed in the resonator hole.
F-MISO was used for PET hypoxic imaging.

A bolus of F-MISO was injected intravenously through a tail vein cannula. 
EPRI acquisition began 1.5 hours post F-MISO injection. 
During EPRI data acquisition,
the trityl spin probe was infused continuously through cannulation at a rate of 240 $\mu$l/h.  
Two or three 7-minute EPRI images were taken in sequence during which
list mode PET data were continuously acquired.
This simultaneous imaging session lasted 20-30 minutes. 
For evaluation, an additional PET-only imaging was conducted beginning 2 hours post F-MISO injection 
to collect PET data free of the interference of the EPRI pulse sequences. 
 
In addition to PET/EPRI imaging, 
MR T2-weighted image of the animal was also obtained for anatomical reference. 
MRI was done separately, after PET/EPRI imaging,
by using a preclinical Bruker 9.4 T scanner (Billerica, MA)
employing the Rapid Imaging with Refocused Echo (RARE) sequence. 
The four non-parallel capillary tubes placed around the animal bed
were used as fiducial markers for registration of EPR and MR images.  
Each capillary tube contained 1 mM of trityl spin probe solution
that is detectable by both EPRI and MRI.

\section{Results}
\subsection{PET/EPRI mutual interference}
\label{subsec:interference}
\begin{figure}[b!]
\begin{center}
\includegraphics[height=8.0cm]{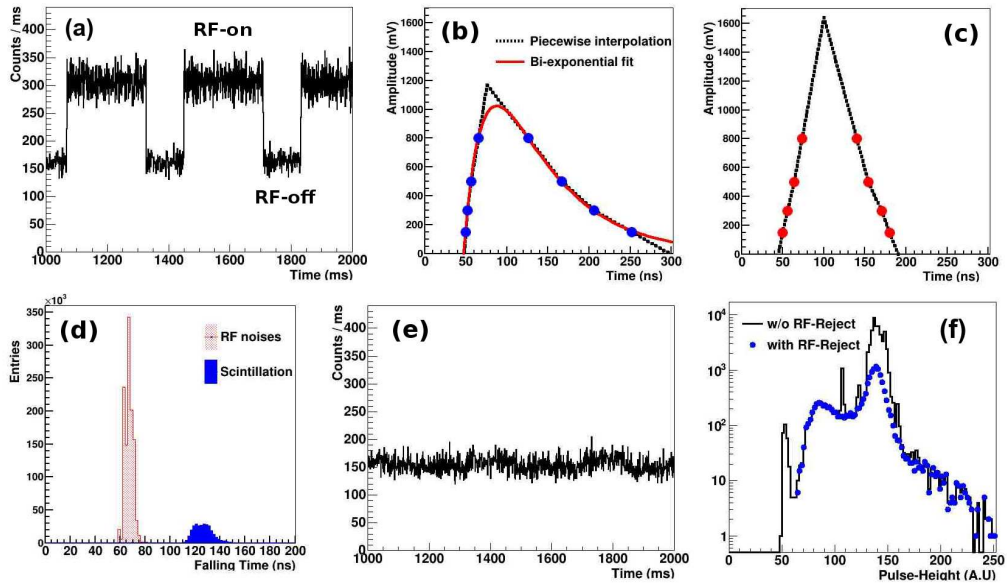} 
\end{center}
\caption{
  (a) PET singles event rate as the EPRI RF pulsing is turned on and off.
  Horizontal axis represents singles event time-stamp recorded in the MVT board.
  (b) A typical scintillation event consisting of 8 MVT samples (blue circles) and
  the signal waveforms estimated from them by using a simple piecewise linear interpolation-based method (dashed-line curve)
  and a bi-exponential fitting function (solid-line curve).
  (c) The waveform of an exemplary spurious RF event picked up by PET data acquisition.
  (d) Histogram of the falling time measured from the data acquired during EPRI RF pulsing.
  (e) Singles event rate after applying the RF rejection.
  (f) Pulse-height histograms for a detector pixel before and after applying the RF rejection.
}
\label{fig:3} 
\end{figure}
Effects on EPRI due to PET was found to be negligible
from the comparison of the capillary tube phantom images obtained before and after installation of the PET subsystem.
However, the PET subsystem, as the detector modules had no RF shielding, 
picked-up EPRI RF pulses during simultaneous PET/EPRI imaging.
To investigate the effect, 
we observed the singles count rate and the pulse shape as determined by the MVT samples. 
Figure~\ref{fig:3}(a) shows that during RF pulsing the singles count rate increases significantly, 
due to the detection of spurious events associated with the EPRI RF pulses.  
However, with MVT data acquisition, 
these spurious RF events can be effectively rejected by off-line data processing.  

As stated above, MVT sampling produces a leading and trailing sample
at each voltage threshold to yield 8 samples for each signal. 
Figures~\ref{fig:3}(b) and \ref{fig:3}(c) show the signal waveforms estimated from these MVT samples (circles)
for a real scintillation pulse and a spurious RF event, respectively.
In this work, a simple piecewise linear interpolation-based method is used for estimating signal waveform.
The leading portion of the waveform between the 1st and the 4th samples is obtained by linearly interpolating these samples,
that before the 1st sample by extrapolating the interpolation line of the 1st and 2nd samples until it reaches zero,
and that after the 4th samples by extrapolating the interpolation line between the 3rd and 4th samples
until it meets the trailing portion of the waveform, which itself is obtained from the 5th to 8th samples in a similar way.
Evidently, the point where the leading and trailing portions meet defines the peak of the waveform.
Also shown in Figure~\ref{fig:3}(b) is the pulse waveform obtained by a bi-exponential fitting
for LYSO/MPPC MVT data that was previously proposed and validated in~\citep{xi_2013}.
Although the piecewise linear interpolation-based method was not as accurate,
it was much faster to compute and was found to yield adequate energy resolution for the proposed system.

It was observed that the spurious RF events have a shorter duration than the scintillation pulses. 
The histogram in Figure~\ref{fig:3}(d) shows that
the falling time (calculated as the time difference between the 5th and 8th MVT samples) 
of the spurious RF events and actual scintillation events have well separated distributions, 
with the former being approximately 67 ns and the latter approximately 130 ns.
Consequently, events having falling time smaller than 95 ns were identified as spurious RF events
and rejected before energy qualification and coincidence filtering. 
Figure~\ref{fig:3}(e) shows the singles count rate after implementation of this rejection method.
Figure~\ref{fig:3}(f) shows the pulse-height histograms of a detector pixel
before and after applying the rejection method on the EPRI-on data.
(Note that the y-axis uses logarithmic scale in Figure~\ref{fig:3}(f).) 
The pulse-height of single event was calculated by integrating the area under the interpolated waveform.

\begin{figure}[h!]
\begin{center}
\includegraphics[width=11.0cm]{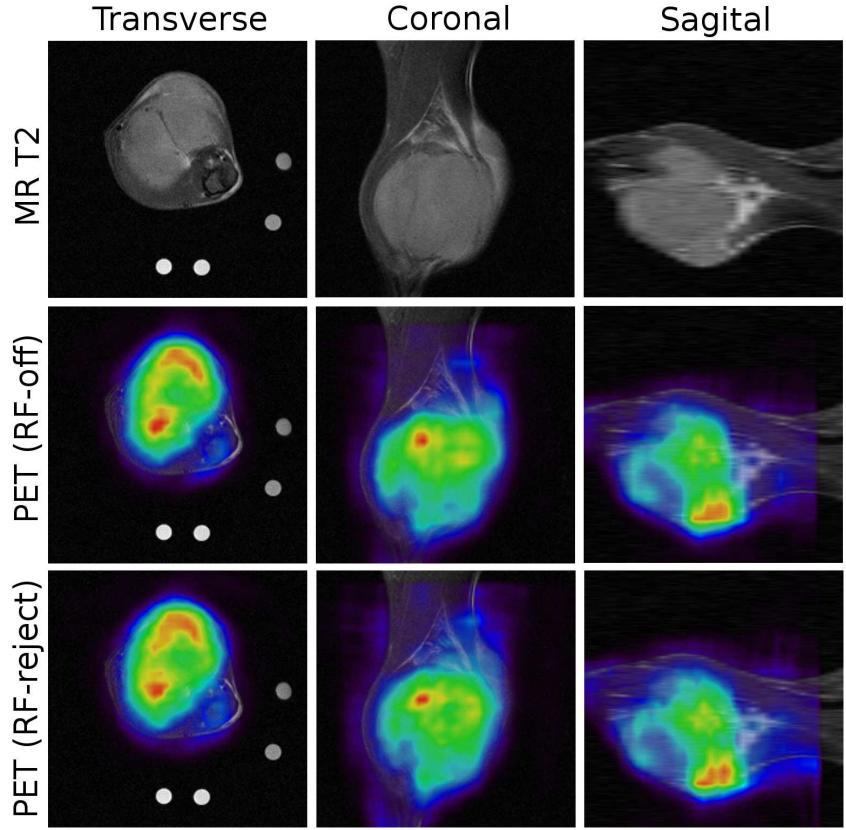} %
\end{center}
\caption{
  (top) A MR T2-weighted image of the tumor in a mouse are acquired for anatomical reference.
  (middle) PET image obtained from the EPRI-off data are registered to the MR image.
  (bottom) PET image for the same slice obtained from the EPRI-on and RF-rejected data.  
}
\label{fig:4} 
\end{figure}

To check the effectiveness of the proposed RF-event rejection method, 
a tumor in the leg of a female mouse was imaged.
4.0 MBq FDG was injected to the female mouse (21.0 gram),
and imaging started 35 minutes post-injection.
Two PET data sets were obtained for comparison: 
one data set was collected while EPRI was off (EPRI-off data), 
and the other set was acquired while EPRI was on but processed 
by employing the RF-event rejection method (EPRI-on, RF-rejected data).
Table~\ref{tab:1} summarizes the acquisition time and the resulting numbers of coincidence events.
The comparable numbers of coincidence events suggests 
the effective removal of spurious RF events by the proposed method.
Figure~\ref{fig:4} shows the PET images obtained from the two data sets, 
co-registered to the MRI T2-weighted image for visualization.
The PET images obtained from the EPRI-on, RF-rejected data set do not show noticeable artifacts,
and are similar to the images from the EPRI-off data.
Note that the data sets were acquired at approximately 15 minutes apart; 
the slight but observable differences between the images possibly reflect temporal physiological changes that occur between the imaging sessions.
\begin{table}[h!]
\begin{center}
\begin{tabular}{|c|c|c|c|}
\hline
 Data set            &  EPR On & EPR On, RF-rejected & EPR Off \\ \hline
 DAQ time (minutes)  &   15.5  & 15.5    & 14.5 \\
 Coincidence events  &  17.4 M     & 0.71 M   & 0.68 M \\ 
\hline
\end{tabular}
\caption{Comparison of the data acquisition time and the number of coincidence event from the two data sets.}
\end{center}
\label{tab:1}
\end{table}
 
\subsection{F-MISO and EPRI pO$_{2}$: animal imaging}
\begin{figure}[h!]
\begin{center}
\includegraphics[width=11.0cm]{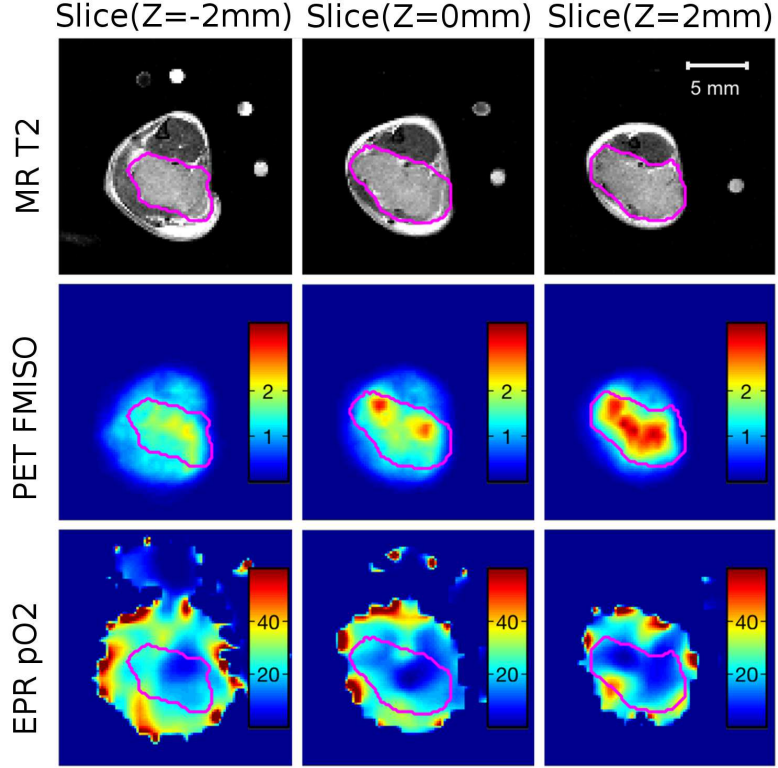} 
\end{center}
\caption{
  (top) A MR T$_2$ image of the tumor in a mouse.
  The tumor boundary is determined based on the MR image and marked by pink color line.
  (middle) A PET image for the same slice shows high F-MISO uptake in the tumor.
  (bottom) pO$_2$ image acquired by the EPRI.
  The hypoxic region within the tumor is indicated by dark blue color. 
}
\label{fig:5} 
\end{figure}

We have conducted 10 mice F-MISO imaging using the PET/EPRI combined system.
A typical image is shown in Figure~\ref{fig:5}
showing a mouse (23.9 gram) with a leg-born SCCVII squamous cell carcinoma
immobilized in way similar to that shown in Figure~\ref{fig:2}(d).
An MR T$_2$ image was acquired for delineating the tumor region 
as an anatomical reference after PET/EPRI imaging.
Figure~\ref{fig:5}(top) shows the tumor boundary marked on the MR images at three different slices.
Figure~\ref{fig:5}(middle) shows the corresponding PET images.
High F-MISO uptake are shown in red and they are observed to appear within the tumor boundary.
Figure~\ref{fig:5}(bottom) shows a pO$_2$ image by the EPRI for the same image slices.
The hypoxic regions within the tumor, defined as pO$_2$ $\leq$ 10 torr, is indicated by dark blue color in the image.
Some hypoxic regions are overlapped in both F-MISO image and EPRI oxygen images. 
However, there is also discrepancy between images.

\section{Discussion}
We have integrated a PET system with an EPRI system and produced successful imaging results for animals. 
Recently, a combined PET/PERI system was reported by Tseytlin et al.~\citep{tseytlin_2018},
showing successful imaging of a multi-modality phantom containing trityl and FDG solution,
but no animal imaging results.
This system is different from ours reported here
in terms of the PET detector technology and the integration of the PET and EPRI systems.
It is a 21 cm inner diameter ring made of 12 detector modules,
each of which consists of a 32x32 array 1.5x1.5x10 mm$^3$ LYSO crystals (1.57 mm pixel)
coupled to a 10x10 array of 3x3 mm$^2$ SiPMs (4.85 mm pitch).
Readout of the 10x10 SiPMs is based on the conventional charge division multiplexing
to reduce the signals to 2x2, and these signals are acquired
by using the conventional analog-to-digital converters (ADC) and TDCs.
As its diameter is larger than the gap between the EPRI magnets,
the PET ring cannot fit with the EPRI system in the way shown in Figure~\ref{fig:1}(c).
Instead, it "sits" on the bottom EPRI magnet with a 20$^{\circ}$ tilt angle
to allow some access to the imaging-sensitive volume.
Another crucial important difference lies in the design of the EPR oxygen imaging system.
The EPR imager of Tseytlin et al. utilizes Rapid Scan (RS) EPR acquisition~\citep{joshi_2005} 
measuring pO$_2$ with spin-spin (SS) relaxation based linewidth measurement
while we use spin-lattice relaxation (SLR) measured with pulse mode inversion recovery acquisition.
SS relaxation based pO$_2$ measurements are more susceptible to confounding broadening of the spin probe linewidth
by the spin probe itself to which SLR is far less susceptible~\citep{epel_2014}.
Thus the proposed system is uniquely suitable to study the effects of hypoxia.

\begin{figure}[h!]
\begin{center}
\includegraphics[height=5.0cm]{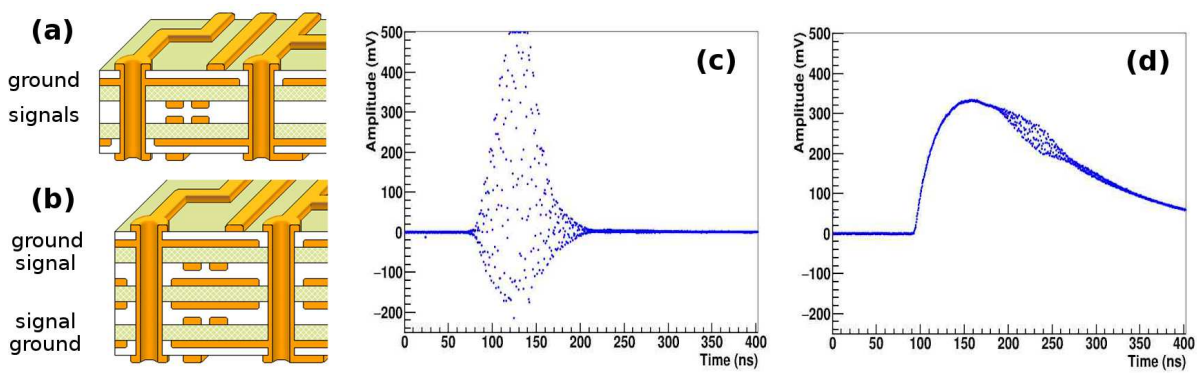} 
\end{center}
\caption{
  (a) The PCB layout of strip-line board currently used in the PET
  (b) The PCB layout in a newly designed strip-line board.
  (c) An exemplary waveform of RF noise using the current PCB.
  (d) An exemplary waveform of RF noise using the new PCB. 
  The RF noise is riding on top of scintillation signal.
}
\label{fig:6} 
\end{figure}

While we reported successful imaging without PET detector shielding,
for optimal imaging performance we believe some shielding is still necessary.
However, we desire to mitigate the shielding requirement as much as possible 
to maintain a simple and compact detector design. 
Recently, we considered a new SLB design to provide inherent shielding of the PET signal traces.
The current SLB design shown in Figure~\ref{fig:6}(a) uses PCB with a single ground plane.
In the new SLB,
another ground plane is added so that   
the SiPM signal traces inside the PCB are laid between the two ground planes,
as shown in Figure~\ref{fig:6}(b).
A preliminary comparison test between the current and new design SLB was conducted 
by using single detector module that was coupled to the current or new SLB.
The result shows that the EPR RF interference is significantly reduced  
with the newly designed SLB.
Exemplary waveforms of RF noise events, recorded by the DRS4 waveform sampler~\citep{ritt_2010},
for the two SLB boards are shown in Figures~\ref{fig:6}(c) and (d).   
The amplitude of RF signal appearing on the PET electronics was reduced from often exceeding 700 mV to 50 mV.
The percentage of spurious RF events in singles 
(without the proposed RF rejection method) also is reduced from 76\% to 2.6\%.
We plan to use this new SLB in our next implementation of the PET system.
We will also consider the use of thin copper foils to wrap the PET insert for improved shielding.

\section{Summary}
Precise delineation of hypoxic region in tumor is essential 
for improving the outcome of oxygen image-guided radiation therapy treatment for hypoxic tumor.
PET imaging with F-MISO has been used for hypoxic tumor imaging in clinical trials. 
However, due to the complexity of F-MISO accumulation
and its possible dependence on factors other than hypoxia, 
F-MISO PET imaging has failed to produce accurate assessment of hypoxic tumor regions {\it in vivo}.
Toward quantitative F-MISO PET imaging, 
we prototyped a PET/EPRI combined system for small animal imaging
for assessing, and potentially calibrating, F-MISO PET hypoxia imaging 
by using simultaneous EPRI oxygen images as reference. 
The initial F-MISO and EPR imaging on mouse model hypoxic tumor 
was successfully conducted using the developed system.

The PET scanner was observed to pick-up RF noises during simultaneous PET/EPRI operation.
An effective RF-event rejection method was developed 
by exploiting the difference between the waveform features of the scintillation and RF events.
A newly designed SLB board that can better shield the PET signals from EPRI pulsing 
is also tested and the result is encouraging.
A more comprehensive animal imaging with the developed PET/EPRI system is undergoing and 
we are in the process of analyzing the correlations
between the resulting F-MISO PET and EPRI oxygen images, 
and with other physiological parameters derived from MRI imaging, 
to develop a more complete understanding of F-MISO uptake by healthy and tumor tissues.

\section*{Acknowledgments}
The authors acknowledge that this work was supported in part by
the National Institutes of Health under Grant numbers R01 CA236385, R01 CA098575, R03 EB027343,
R01 EB029948, R50 CA211408, P30 CA014599, P41 EB002034, S10 OD025265,  and T32 EB002103. 
We are also grateful for excellent support from Integrated Small Animal Imaging Research Resource
and the Cyclotron facility at the University of Chicago.

\bibliographystyle{JHEP}

\bibliography{petepr}

\end{document}